\documentclass[10 pt, letter]{article}
\usepackage{graphicx}
\usepackage{caption}
\usepackage{amsmath}
\usepackage{amsthm}
\usepackage{enumerate}
\usepackage{subfig}
\usepackage{float}
\usepackage{amsfonts}
\usepackage{amssymb}
\usepackage{anysize}
\usepackage{helvet}
\usepackage{setspace}
\usepackage{multirow}
\usepackage{tabularx}
\usepackage{tabulary}
\usepackage{longtable}
\usepackage{float}
\usepackage{mathtools}
\begin{document}

\title{\bf About Factorization of Quantum States with Few Qubits}
\author{G.V. L\'opez$^{(1)}$\footnote{gulopez@udgserv.cencar.udg.mx}~, G. Montes$^{(1)}$, M. Avila$^{(2)}$, and J. Rueda-Paz$^{(2)}$\\
\small \emph{$^{(1)}$Departamento de F\'{i}sica, Universidad de Guadalajara,}\\
\small \emph{Blvd. Marcelino Garc\'{i}a Barragan y Calzada Ol\'{i}mpica, 44200 Guadalajara, Jalisco, Mexico}\\Ê\\
\small\emph{$^{(2)}$Centro Univeristario UAEM Valle de Chalco,UAEMex,}\\
\small\emph{Mar\'{\i}a Isabel, C.P. 56615, Valle de Chalco,}\\
\small\emph{Estado de M\'exico, M\'exico.}}
\date{\today}
\maketitle
%
\begin{abstract}
\noindent
We study the factorization conditions of a wave function made up  of states of two, three and four qubits  and propose and analytical expression which can characterize entangled 
states in terms of the coefficients of the wave function and density matrix elements.
\end{abstract}

\centerline{{\bf PACS:} 03.67.-a, 03.67.Hk, 03.67.Mn, 03.67.Lx, 03.65.Yz }
\newpage
\section{ Introduction}
In quantum mechanics, multiple dimensional or multiple particles systems are characterized by the tensor product of the Hilbert subspaces \cite{Sakurai}, where each subspace is associated to each element. It is well known \cite{Mess} that when there is not interaction among these elements, the wave  function is just the the tensor product of the wave functions of each element, that is, the tensor product of the wave function associated to each element determines the non interacting characteristic of the elements in a quantum system. If interaction occurs at some time among these elements , this tensor product disappears, and the wave function becomes entangled \cite{Sch}. So, it is necessary to point out that if the wave function is not factorized , the wave function is entangled. In this way, the characterization of factorization is somewhat equivalent  to the characterization of entanglement.  In this paper, we will follow this line of ideas to determine  the characterization of an entangled state \cite{Horo,Woott,Seev, Aoli, Pope, Love, Zhi, Huber}.
\section{Factorized State}
When a quantum systems is made up of several quantum subsystems, where the ith-subsystem  is characterized by a Hamiltonian $H_i$ and a Hilbert space ${\cal E}_i$, corresponding a two-states  and having a basis $\{|\xi_i\rangle\}_{\xi_i=0,1}$, the Hilbert space is written as the tensorial product of each subsystem, ${\cal E}={\cal E}_n\otimes\dots\otimes{\cal E}_1$ (for n-subsystems), The state in each subsystems is defined as a qubit in quantum computation and information theory \cite{Niel_QC} and is given by
\begin{equation}\label{b1}
|\psi_i\rangle=a_i|0\rangle+b_i|1\rangle, \quad\quad |a_i|^2+|b_i|^2=1,
\end{equation}
where $\{|0\rangle, |1\rangle\}$ is the basis of the two-states Hilbert subspace ${\cal E}_i$.  A general state $|\Psi\rangle$  in the Hilbert space ${\cal E}$ can be written as
\begin{equation}
|\Psi\rangle=\sum_{\xi}C_{\xi}|\xi\rangle, \quad\quad\hbox{with}\quad \sum_{\xi}|C_{\xi}|^2=1, 
\end{equation}
where $C_{\xi}'s$ are complex numbers, and $|\xi\rangle$ is an element of the basis of ${\cal E}$,
\begin{equation}
|\xi\rangle=|\xi_n\dots\xi_1\rangle=|\xi_n\rangle\otimes\dots\otimes|\xi_1\rangle,\quad\quad \xi_k=0,1\quad k=1,\dots,n.
\end{equation}
A full factorized state in this space  is 
\begin{equation}
|\Psi\rangle=|\psi_n\rangle\otimes\dots\otimes|\psi_1\rangle,
\end{equation}
where $|\psi_k\rangle$ is given by (\ref{b1}). \\ \\
{\bf For $\bf n=2$}, one has a general state  $|\Psi\rangle\in{\cal E}$,
\begin{equation}\label{eq1}
|\Psi\rangle=C_1|00\rangle+C_2|01\rangle+C_3|10\rangle+C_4|11\rangle.
\end{equation}
Let us assume that this state can be written as 
\begin{equation}\label{f1}
|\Psi\rangle=|\psi_2\rangle\otimes|\psi_1\rangle,
\end{equation}
with $|\psi_k\rangle,k=1,2$ given by (\ref{b1}). Then, substituting (\ref{b1}) in (\ref{f1})and equaling coefficients with (\ref{eq1}), it follows that
\begin{equation}
C_1=a_1a_2,\quad C_2=a_1b_2,\quad C_3=b_1a_2,\quad C_4=b_1b_2.
\end{equation}
From these expression one obtains a single condition for factorization,
\begin{equation}
C_1C_4=C_2C_3.
\end{equation}
Thus, if this condition is not satisfied, the state (\ref{eq1}) represents an entangled state. So, one can use the following known expression \cite{Alb} as a characterization of an entangled state
\begin{equation}
C_{\Psi}^{(2)}=2|C_1C_4-C_2C_3|.
\end{equation}
{\bf For $\bf n=3$}, a general state in the Hilbert space ${\cal E}$ is
\begin{equation}\label{eq2}
|\Psi\rangle=C_1|00\rangle+C_2|001\rangle+C_3|010\rangle+C_4|011\rangle+C_5|100\rangle+C_6|101\rangle+C_7|110\rangle+C_8|111\rangle.
\end{equation}
Assuming that this wave function can be written as $|\Psi\rangle=\psi_3\rangle\otimes|\psi_2\rangle\otimes|\psi_1\rangle$ with $\psi_k\rangle$ given by (\ref{b1}), and after some identifications  (as before) and rearrangements, one gets
\begin{subequations}
\begin{eqnarray}
&& \frac{C_1}{C_2}=\frac{C_3}{C_4}=\frac{C_5}{C_6}=\frac{C_7}{C_8}=\frac{a_3}{b_3},\quad\quad b_3\not=0\\ \nonumber \\
&& \frac{C_1}{C_3}=\frac{C_2}{C_4}=\frac{C_5}{C_7}=\frac{C_6}{C_8}=\frac{a_2}{b_2},\quad\quad b_2\not=0\\ \nonumber \\
&& \frac{C_1}{C_5}=\frac{C_2}{C_6}=\frac{C_3}{C_7}=\frac{C_4}{C_8}=\frac{a_1}{b_1},\quad\quad b_1\not=0
\end{eqnarray} 
\end{subequations}
These expression reflex a parallelism between the complex vectors $(C_1,C_3,C_5,C_7)$ and $(C_2,C_4,C_6,C_8)$, the vectors $(C_1,C_2,C_5, C_6)$ and $(C_3,C_4,C_7,C_8)$, and the vectors $(C_1,C_2,C_3,C_4)$ and $(C_5,C_6,C_7,C_8)$. In addition, they bring about the following eight independent relations
\begin{eqnarray}
& &C_1C_4-C_2C_3=0,\quad C_1C_6-C_2C_5=0,\quad C_1C_8-C_2C_7=0,\quad C_3C_6-C_4C_5=0\\
& &C_3C_8-C_4C_7=0,\quad C_5C_8-C_6C_7=0,\quad C_1C_7-C_3C_5=0,\quad C_2C_8-C_4C_6=0.
\end{eqnarray}
If one of these expression is not satisfied, the wave function (\ref{eq2}) represents an entangled state. Therefore, one can propose the following expression to characterize an entangled state
\begin{eqnarray}\label{c3Cs}
C_{\Psi}^{(3)}&=2|C_1C_4-C_2C_3|+2|C_1C_6-C_2C_5|+2|C_1C_8-C_2C_7|+2|C_3C_6-C_4C_5|\nonumber\\ 
&\quad+2|C_3C_8-C_4C_7|+2|C_5C_8-C_6C_7|+2|C_1C_7-C_3C_5|+2|C_2C_8-C_4C_6|.
\end{eqnarray}
{\bf For $\bf n=4$}, a general state in the Hilbert space ${\cal E}$ is of the form
\begin{eqnarray}\label{eq3}
|\Psi\rangle&=C_1|0000\rangle+C_2|0001\rangle+C_3|010\rangle+C_4|0011\rangle\nonumber\\
&\quad+C_5|0100\rangle+C_6|0101\rangle+C_7|0110\rangle+C_8|0111\rangle\nonumber\\
&\quad+C_9|100\rangle+C_{10}|1001\rangle+C_{11}|110\rangle+C_{12}|1011\rangle\nonumber\\
&\quad+C_{13}|1100\rangle+C_{14}|1101\rangle+C_{15}|1110\rangle+C_{16}|1111\rangle.
\end{eqnarray}
Assuming this function can be expressed as $|\Psi\rangle=|\psi_4\rangle\otimes|\psi_3\rangle\otimes|\psi_2\rangle\otimes|\psi_1\rangle$ with $|\psi_k\rangle, k=1,2,3,4$ given by (\ref{b1}), and after some identifications and rearrangements, one gets
\begin{eqnarray*}
&\frac{C_1}{C_2}=\frac{C_3}{C_4}=\frac{C_5}{C_6}=\frac{C_7}{C_8}=\frac{C_9}{C_{10}}=\frac{C_{11}}{C_{12}}=\frac{C_{13}}{C_{14}}=\frac{C_{15}}{C_{16}}\\ \nonumber\\
&\frac{C_1}{C_3}=\frac{C_2}{C_4}=\frac{C_5}{C_7}=\frac{C_6}{C_8}=\frac{C_9}{C_{11}}=\frac{C_{10}}{C_{12}}=\frac{C_{13}}{C_{15}}=\frac{C_{14}}{C_{16}}\\ \nonumber\\
&\frac{C_1}{C_5}=\frac{C_2}{C_6}=\frac{C_3}{C_7}=\frac{C_4}{C_8}=\frac{C_9}{C_{13}}=\frac{C_{10}}{C_{14}}=\frac{C_{11}}{C_{15}}=\frac{C_{12}}{C_{16}}\\ \nonumber\\
&\frac{C_1}{C_9}=\frac{C_2}{C_{10}}=\frac{C_3}{C_{11}}=\frac{C_4}{C_{12}}=\frac{C_5}{C_{13}}=\frac{C_6}{C_{14}}=\frac{C_7}{C_{15}}=\frac{C_8}{C_{16}},
\end{eqnarray*}
expressing similar parallelism we mentioned before. Each row gives us 28 relations, having a total of 112 possible relations, and from these relations, one can get the following 36 independent conditions
\begin{subequations}
\begin{eqnarray}
&C_1C_4-C_2C_3=0\quad\quad          C_4C_{13}-C_7C_{10}=0\quad\quad C_1C_6-C_2C_5=0\quad\quad  C_4C_{14}-C_6C_{12}=0\\
&C_1C_8-C_3C_6=0\quad\quad          C_4C_{15}-C_3C_{16}=0\quad\quad C_1C_{10}-C_2C_9=0\quad\quad     C_4C_{16}-C_8C_{12}=0\\
&C_1C_{11}-C_3C_9=0\quad\quad      C_5C_8-C_6C_7=0\quad\quad C_1C_{12}-C_2C_{11}=0\quad\quad C_5C_{14}-C_6C_{13}=0\\
&C_1C_{14}-C_9C_6=0\quad\quad       C_5C_{15}-C_7C_{13}=0\quad\quad C_1C_{15}-C_5C_{11}=0\quad\quad  C_5C_{16}-C_7C_{14}=0\\
&C_2C_8-C_4C_6=0\quad\quad            C_6C_{11}-C_5C_{12}=0\quad\quad C_2C_{12}-C_4C_{10}=0\quad\quad   C_6C_{15}-C_{16}C_8=0\\
&C_2C_{13}-C_5C_{10}=0\quad\quad   C_6C_{16}-C_8C_{14}=0\quad\quad C_2C_{14}-C_6C_{10}=0\quad\quad   C_7C_{16}-C_8C_{15}=0\\
&C_2C_{16}-C_{10}C_8=0\quad\quad   C_7C_{12}-C_8C_{11}=0\quad\quad C_3C_8-C_4C_7=0\quad\quad             C_9C_{12}-C_{10}C_{11}=0\\
&C_3C_{15}-C_7C_{11}=0\quad\quad   C_9C_{14}-C_{10}C_{11}=0\quad\quad C_3C_{13}-C_{11}C_5=0\quad\quad   C_9C_{15}-C_{11}C_{13}=0\\
&C_{10}C_{16}-C_{12}C_{14}=0\quad\quad  C_{11}C_{16}-C_{12}C_{15}=0\quad\quad C_{10}C_{15}-C_{11}C_{14}=0\quad  C_{13}C_{16}-C_{14}C_{15}=0.
\end{eqnarray}
\end{subequations}
Again, if one of these expression fail to happen, (\ref{eq3}) will represents an entangled state. Thus, one can propose the following expression to characterize an entangled state made up of 4-qubits basis
\begin{eqnarray}
C_{\Psi}^{(4)}&=&2|C_1C_4-C_2C_3|+ 2| C_4C_{13}-C_7C_{10}|+2|C_1C_6-C_2C_5|+2|C_4C_{14}-C_6C_{12}|\nonumber\\
&&+2|C_1C_8-C_3C_6|+2|C_4C_{15}-C_3C_{16}|+2|C_1C_{10}-C_2C_9|+2|C_4C_{16}-C_8C_{12}|\nonumber\\
&&+2|C_1C_{11}-C_3C_9|+2|C_5C_8-C_6C_7|+2|C_1C_{12}-C_2C_{11}|+2|C_5C_{14}-C_6C_{13}|\nonumber\\
&&+2|C_1C_{14}-C_9C_6|+2|C_5C_{15}-C_7C_{13}|+2|C_1C_{15}-C_5C_{11}|+2|C_5C_{16}-C_7C_{14}|\nonumber\\
&&+2|C_2C_8-C_4C_6|+2|C_6C_{11}-C_5C_{11}|+2|C_2C_{12}-C_4C_{10}|+2|C_6C_{15}-C_{16}C_8|\nonumber\\
&&+2|C_2C_{13}-C_5C_{10}|+2|C_6C_{16}-C_8C_{14}|+2|C_2C_{14}-C_6C_{10}|+2|C_7C_{16}-C_8C_{15}|\nonumber\\
&&+2|C_2C_{16}-C_{10}C_8|+2|C_7C_{12}-C_8C_{11}|+2|C_3C_8-C_4C_7|+2|C_9C_{12}-C_{10}C_{11}|\nonumber\\
&&+2|C_3C_{15}-C_7C_{11}|+2| C_9C_{14}-C_{10}C_{11}|+2|C_3C_{13}-C_{11}C_5|+2|C_9C_{15}-C_{11}C_{13}|\nonumber\\
&&+2|C_{10}C_{16}-C_{12}C_{14}|+2|C_{11}C_{16}-C_{12}C_{15}|+2|C_{10}C_{15}-C_{11}C_{14}|+2|C_{13}C_{16}-C_{14}C_{15}|.\nonumber\\
\end{eqnarray}
\newpage
As we can see from these examples, the number of conditions needed to characterize a factorized state (or entangled state) grows exponentially with the number of qubits. So, characterization of an entangled state for n-qubits in general becomes a very hard work. Now,  in terms of the density matrix elements, one could have the characterization of entangled states made up of 2,3, and 4 qubits as
\begin{eqnarray}
& C^{(2)}_{\rho}=2\sqrt{\rho_{11}\rho_{44}+\rho_{22}\rho_{33}-2Re(\rho_{12}\rho_{43}) }\\ \nonumber\\
& C^{(3)}_{\rho}=2\sqrt{\rho_{11}\rho_{44}+\rho_{22}\rho_{33}-2Re(\rho_{12}\rho_{43}) }
+2\sqrt{\rho_{11}\rho_{66}+\rho_{22}\rho_{55}-2Re(\rho_{12}\rho_{65}) }\nonumber\\
&\quad+2\sqrt{\rho_{11}\rho_{88}+\rho_{22}\rho_{77}-2Re(\rho_{12}\rho_{87}) }
+2\sqrt{\rho_{33}\rho_{66}+\rho_{44}\rho_{55}-2Re(\rho_{34}\rho_{65}) }\nonumber\\
&+2\sqrt{\rho_{33}\rho_{88}+\rho_{44}\rho_{77}-2Re(\rho_{34}\rho_{87}) }
+2\sqrt{\rho_{55}\rho_{88}+\rho_{66}\rho_{77}-2Re(\rho_{56}\rho_{87}) }\nonumber\\
&\quad+2\sqrt{\rho_{11}\rho_{77}+\rho_{33}\rho_{55}-2Re(\rho_{13}\rho_{75}) }
+2\sqrt{\rho_{22}\rho_{88}+\rho_{44}\rho_{66}-2Re(\rho_{24}\rho_{86}) }.\label{c3rho}\\ \nonumber\\
&C_{\rho}^{(4)}=2\sqrt{\rho_{11}\rho_{44}+\rho_{22}\rho_{33}-2Re(\rho_{12}\rho_{43})Ê}+ 2\sqrt{ \rho_{44}\rho_{13,13}+\rho_{77}\rho_{10,10}-2Re(\rho_{47}\rho_{13,10}) }\nonumber\\
&\quad+2\sqrt{\rho_{11}\rho_{66}+\rho_{22}\rho_{55}-2Re(\rho_{12}\rho_{65}) }+2\sqrt{\rho_{44}\rho_{14,14}+\rho_{66}\rho_{12,12}-2Re(\rho_{46}\rho_{14,12}) }\nonumber\\
&\quad+2\sqrt{\rho_{11}\rho_{88}+\rho_{33}\rho_{66}-2Re(\rho_{13}\rho_{86}) }+2\sqrt{\rho_{44}\rho_{15,15}+\rho_{33}\rho_{16,16}-2Re(\rho_{43}\rho_{15,16}) }|\nonumber\\
&\quad+2\sqrt{\rho_{11}\rho_{10,10}+\rho_{22}\rho_{99}-2Re(\rho_{12}\rho_{10,9}) }+2\sqrt{\rho_{44}\rho_{16,16}+\rho_{88}\rho_{12,12}-2Re(\rho_{48}\rho_{16,12}) }\nonumber\\
&\quad+2\sqrt{\rho_{11}\rho_{11,11}+\rho_{33}\rho_{99}-2Re(\rho_{13}\rho_{11,9}) }+2\sqrt{\rho_{55}\rho_{88}+\rho_{66}\rho_{77}-2Re(\rho_{56}\rho_{87}) }\nonumber\\
&\quad+2\sqrt{\rho_{11}\rho_{12,12}+\rho_{22}\rho_{11,11}-2Re(\rho_{12}\rho_{12,11}) }+2\sqrt{\rho_{55}\rho_{14,14}+\rho_{66}\rho_{13,13}-2Re(\rho_{56}\rho_{14,13}) }\nonumber\\
&\quad+2\sqrt{\rho_{11}\rho_{14,14}+\rho_{99}\rho_{66}-2Re(\rho_{19}\rho_{14,6}) }+2\sqrt{\rho_{55}\rho_{15,15}+\rho_{77}\rho_{13,13}-2Re(\rho_{57}\rho_{15,13}) }\nonumber\\
&\quad+2\sqrt{\rho_{11}\rho_{15,15}+\rho_{55}\rho_{11,11}-2Re(\rho_{15}\rho_{15,11}) }+2\sqrt{\rho_{55}\rho_{16,16}+\rho_{77}\rho_{14,14}-2Re(\rho_{57}\rho_{16,14}) }\nonumber\\
&\quad+2\sqrt{\rho_{22}\rho_{88}+\rho_{44}\rho_{66}-2Re(\rho_{24}\rho_{86}) }+2\sqrt{\rho_{66}\rho_{11,11}+\rho_{55}\rho_{12,12}-2Re(\rho_{6,5}\rho_{11,12}) }\nonumber\\
&\quad+2\sqrt{\rho_{22}\rho_{12,12}+\rho_{44}\rho_{10,10}-2Re(\rho_{24}\rho_{12,10}) }+2\sqrt{\rho_{66}\rho_{15,15}+\rho_{16,16}\rho_{88}-2Re(\rho_{6,16}\rho_{15,8}) }\nonumber\\
&\quad+2\sqrt{\rho_{22}\rho_{13,13}+\rho_{55}\rho_{10,10}-2Re(\rho_{25}\rho_{13,10}) }+2\sqrt{\rho_{66}\rho_{16,16}+\rho_{88}\rho_{14,14}-2Re(\rho_{68}\rho_{16,14}) }\nonumber\\
&\quad+2\sqrt{\rho_{22}\rho_{14,14}+\rho_{66}\rho_{10,10}-2Re(\rho_{2,6}\rho_{14,10}) }+2\sqrt{\rho_{77}\rho_{16,16}+\rho_{88}\rho_{15,15}-2Re(\rho_{78}\rho_{16,15}) }\nonumber\\
&\quad+2\sqrt{\rho_{22}\rho_{16,16}+\rho_{10,10}\rho_{88}-2Re(\rho_{2,10}\rho_{16,8}) }+2\sqrt{\rho_{77}\rho_{12,12}+\rho_{88}\rho_{11,11}-2Re(\rho_{78}\rho_{12,11}) }\nonumber\\
&\quad+2\sqrt{\rho_{33}\rho_{88}+\rho_{44}\rho_{77}-2Re(\rho_{34}\rho_{87}) }+2\sqrt{\rho_{99}\rho_{12,12}+\rho_{10,10}\rho_{11,11}-2Re(\rho_{9,10}\rho_{12,11}) }\nonumber\\
&\quad+2\sqrt{\rho_{33}\rho_{15,15}+\rho_{77}\rho_{11,11}-2Re(\rho_{37}\rho_{15,11}) }+2\sqrt{ \rho_{99}\rho_{14,14}+\rho_{10,10}\rho_{11,11}-2Re(\rho_{9,10}\rho_{14,11}) }\nonumber\\
&\quad+2\sqrt{\rho_{33}\rho_{13,13}+\rho_{11,11}\rho_{55}-2Re(\rho_{3,11}\rho_{13,5}) }+2\sqrt{\rho_{99}\rho_{15,15}+\rho_{11,11}\rho_{13,13}-2Re(\rho_{9,11}\rho_{1513}) }\nonumber\\
&\quad+2\sqrt{\rho_{10,10}\rho_{16,16}+\rho_{12,12}\rho_{14,14}-2Re(\rho_{10,12}\rho_{1614}) }+2\sqrt{ \rho_{11,11}\rho_{16,16}+\rho_{12,12}\rho_{15,15}-2Re(\rho_{11,12}\rho_{16,15}) }\nonumber\\
&\quad+2\sqrt{\rho_{10,10}\rho_{15,15}+\rho_{11,11}\rho_{14,14}-2Re(\rho_{10,11}\rho_{15,14}) }+2\sqrt{\rho_{13,13}\rho_{16,16}+\rho_{14,14}\rho_{15,15}-2Re(\rho_{13,14}\rho_{16,15}) }.\nonumber\\
\end{eqnarray}
For other considerations of entanglement multiqubit entanglement see \cite{Seev, Zhi, Ry, Dur}. 
Figure 1 below shows the values of the expressions (\ref{c3Cs}) and (\ref{c3rho}) for 50 entangled states made up of 3-qubits basis and with values $C_j$, with $j=1,\dots,6$  randomly generated. As we can see, the values obtained with the coefficients $C_j's$ and with the density matrix elements $\rho_{nm}$ are the same.  Figure 2 shows for the $|W\rangle$ state,
\begin{equation}
|W\rangle=C_2|001\rangle+C_3|010\rangle+C_5|100\rangle, \quad\quad |C_2|^2+|C_3|^2+|C_5|^2=1,
\end{equation}  
the possible values of $C^{(3)}(\Psi)$. As we can see, there are four possible maxima corresponding to the values $C_2=C_3=C_5=\pm 1/\sqrt{3}$ and four maxima corresponding to the values $C_5=0$,  $C_2=C_3=\pm 1/\sqrt{2}$, related to semi-factorized state $|0\rangle\otimes(C_2|01\rangle+C_3|10\rangle)$.  Figure 3 shows the possible values of $C^{(3)}(\Psi)$ for the state
\begin{equation}
|GHZ\rangle= C_1|000\rangle + C_8|111\rangle, \quad\quad |C_1|^2+|C_8|^2=1.
\end{equation}
The maximum value of $C^{(3)}(\Psi)$ is gotten for $C_1=C_8=1/\sqrt{2}$, as one would expect. 
\begin{figure}[H]
\includegraphics[scale=0.60,angle=0]{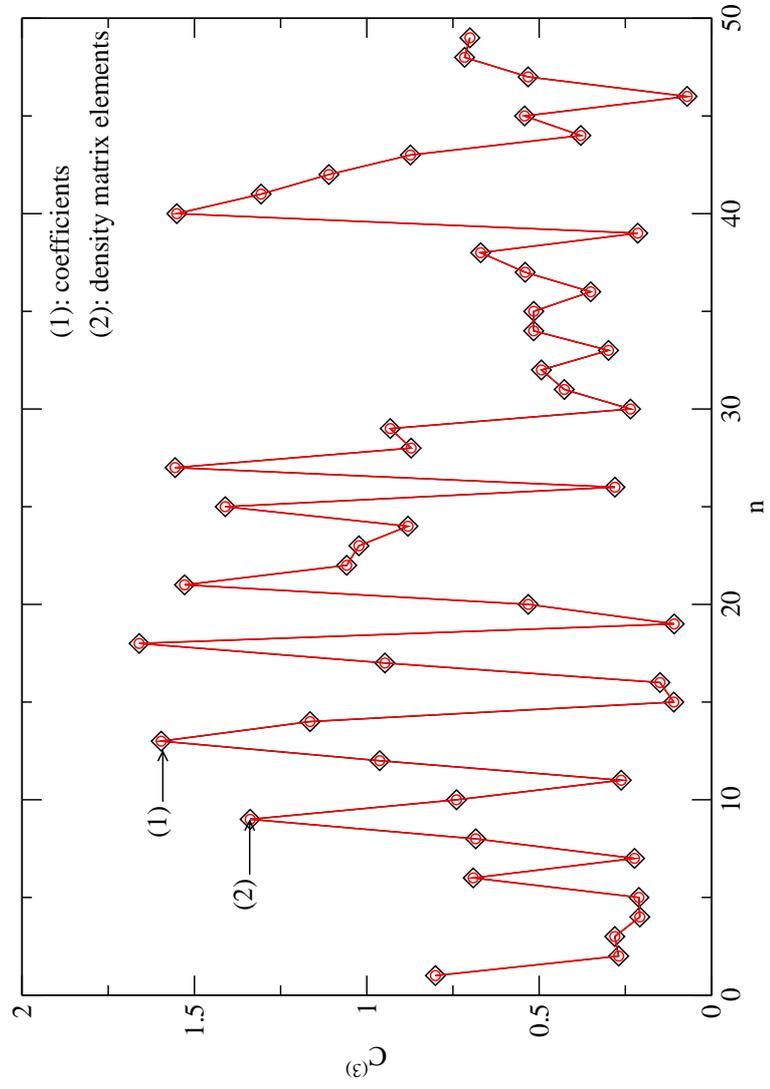}
\centering
    \caption{ $C^{(3)}$ for arbitrary entangled state. }
\end{figure}
\begin{figure}[H]
\includegraphics[scale=0.75,angle=0]{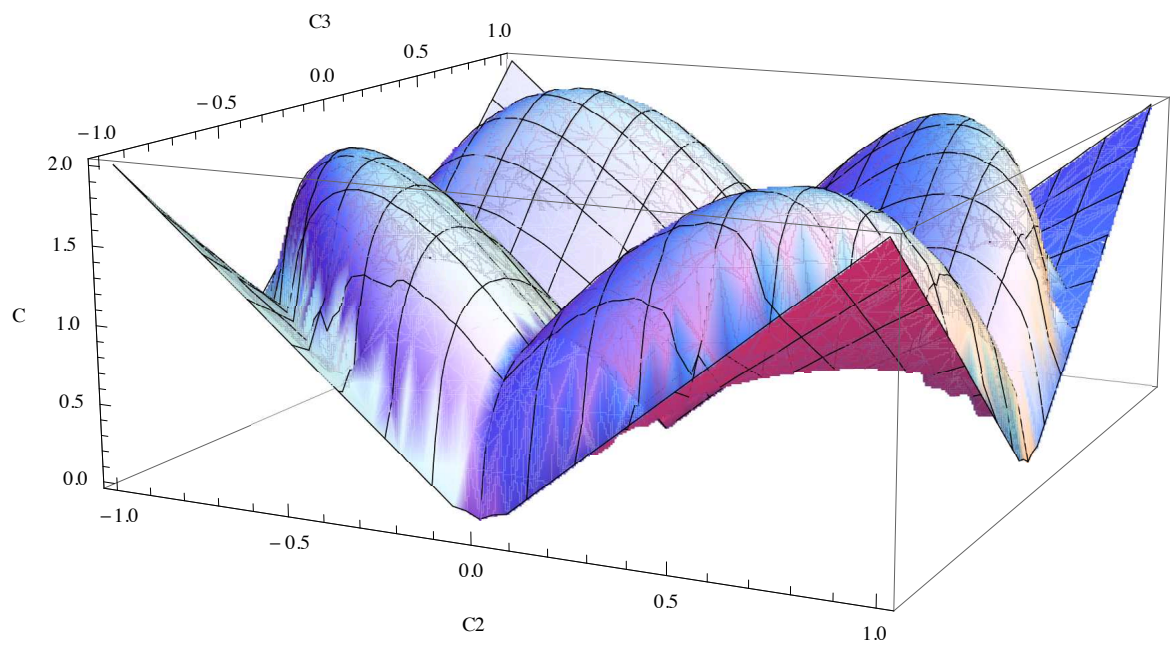}
\centering
    \caption{ $C^{(3)}$ for the state $|W\rangle=C_1|001\rangle+C_2|010\rangle+C_3|100\rangle$ such that \\
    $|C_1|^2+|C_2|^2+|C_3|^2=1$. }
\end{figure}
\begin{figure}[H]
\includegraphics[scale=0.60,angle=0]{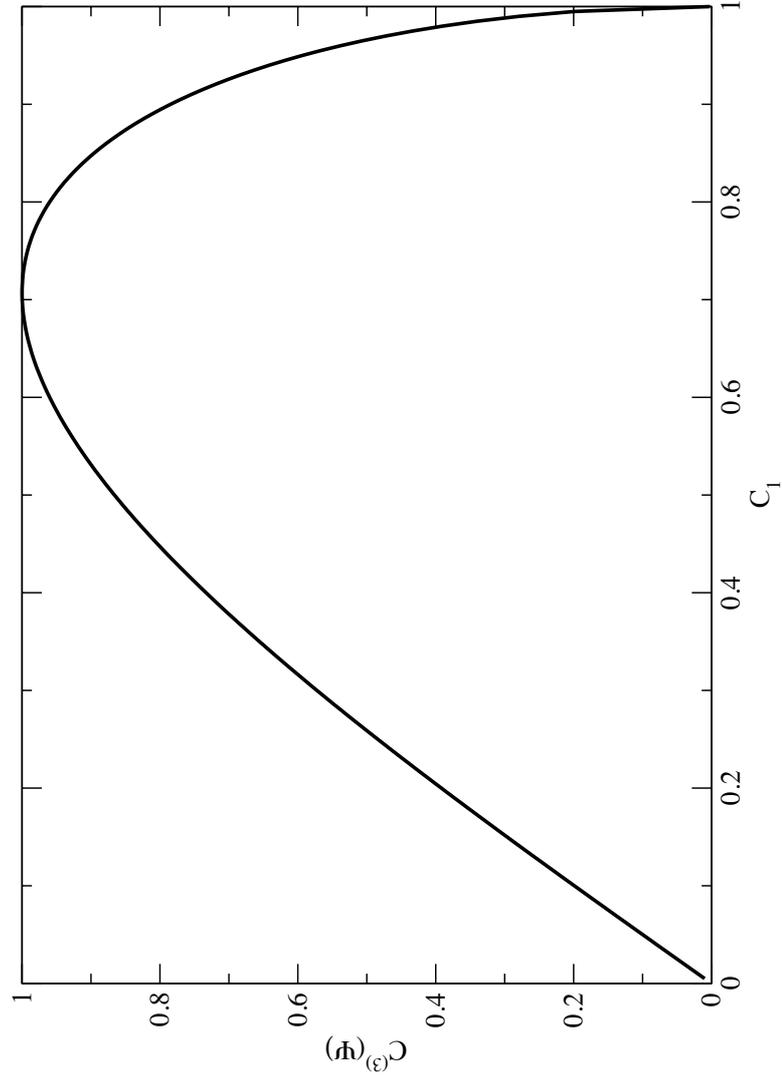}
\centering
    \caption{ $C^{(3)}$ for the state $|GHZ\rangle=C_1|000\rangle+C_8|111\rangle$ such that   $|C_1|^2+|C_8|^2=1$. }
\end{figure}
For a Hilbert space ${\cal E}$ generated by n-qubits, $C^{(n)}$ defines a continuous function $C^{(n)}:{\cal E}\to \Re^+$ with $\Re^+=[0,+\infty)$, $c\ge 1$. Since the coefficients of the wave function defines a compact set on the real space $\Re^{2n}$ due to the relation $\sum_i |C_i|^2=1$, the image of this compact set is a compact set in $\Re^+$ \cite{Ko}. Thus, a normalization factor is possible to introduce on this function to define any  compact set $[0,c]$, which it is not important. 
\newpage
\section{Dynamical consideration}
Following Lloyd's idea \cite{Lloyd_93}, consider a linear chain of nuclear spin one half, separated by some distance and inside a magnetic in a direction $z$, ${\bf B}(z)=(0,0,B_0(z))$,  and making and angle $\theta$ with respect  this linear chain. Choosing this angle such that $\cos\theta=1/\sqrt{3}$, the dipole-dipole interaction is canceled, the Larmore's frequency for each spin is different, $\omega_k=\gamma B_0(z_k)$ with 
$\gamma$  the gyromagnetic ratio. The magnetic moment of the nucleus $\vec\mu_k$ is related with its spin through the relation $\vec\mu_k=\gamma {\bf S}_k$, and the interaction energy between the magnetic field and magnetic moments is $H_{int}=-\sum_k\vec\mu_k\cdot{\bf B}(z_k)=-\sum_k\omega_kS_k^z$. If in addition, one has first and second neighbor Ising interaction, the Hamiltonian of the system is just \cite{Lopez1,Berman}
\begin{equation}
H_s=-\sum_{k=1}^N\omega_kS_k^z-\frac{2J}{\hbar}\sum_{k=1}^{N-1}S_k^zS_{k+1}^z-\frac{2J'}{\hbar}\sum_{k=1}^{N-2}S_k^zS_{k+2}^z,
\end{equation} 
where $N$ is the number of nuclear spins in the chain (or qubits), $J$ and $J'$ are the coupling constant of the nucleus at first and second neighbor. Using the basis of the register of N-qubits, $\{|\xi_{_N},\dots,\xi_1\rangle\}$ with $\xi_k=0,1$, one has that $S_k^z|\xi_k\rangle=(-1)^{\xi_k}\hbar|\xi_k\rangle/2$. Therefore, the Hamiltonian is diagonal on this basis, and its eigenvalues are
\begin{equation}
E_{\xi}=-\frac{\hbar}{2}\sum_{k=1}^N(-1)^{\xi_k}\omega_k-\frac{J\hbar}{2}\sum_{k=1}^{N-1}(-1)^{\xi_k+\xi_{k+1}}-\frac{J'\hbar}{2}\sum_{k=1}^{N-2}(-1)^{\xi_k+\xi_{k+2}}.
\end{equation}
Consider now that the environment is characterized by a Hamiltonian $H_e$ and its interacting with the quantum system with Hamiltonian $H_s$. Thus,  the total Hamiltonian would be $H=H_s+H_e+H_{se}$, where $H_{se}$ is the part of the Hamiltonian which takes into account the interaction system-environment, and the equation one would need to solve, in terms of the density matrix, is ~\cite{Fano,von}
\begin{equation}
i\hbar\frac{\partial \rho_t}{\partial t}= [H,\rho_t],
\end{equation} 
where $\rho_t=\rho_t(s,e)$ is the density matrix which depends on the system and environment coordinates. The evolution of the system is unitary, but it is not possible to solve this equation due to a lot of degree of freedom. Therefore, under some approximations and tracing over the environment coordinates ~\cite{Davies}, it is possible to arrive to a Lindblad type of equation ~\cite{Bre_OQS,Alicki} for the reduced density matrix $\rho(s)=tr_e(\rho_t)$,
\begin{equation}
i\hbar\frac{\partial \rho}{\partial t} = [H_s,\rho] + \sum_{i=1}^{I}\biggl\{V_i\rho V_{i}^{\dagger} - \frac{1}{2}V_{i}^{\dagger}V_i\rho - \frac{1}{2} \rho V_{i}V_{i}^{\dagger}\biggr\}
\end{equation}
where  $V_i$ are called Kraus' operators. This equation is not unitary and Markovian (without memory of the dynamical process). This equation can be written in the interaction picture, through the transformation
 $    \tilde{\rho} = U\rho U^\dagger$ with $ U = e^{iH_st/\hbar}$,  
 as
\begin{equation}\label{LinU}
  i\hbar \frac{\partial\tilde \rho}{\partial t} = \tilde{\mathcal{L}}(\tilde\rho),
\end{equation}
where $\tilde{\mathcal{L}}(\tilde\rho)$ is the Lindblad  operator
\begin{equation}
\tilde{\mathcal{L}}(\tilde\rho)= \sum_{i=1}^{I} \biggl\{\tilde V_i\tilde\rho \tilde V_{i}^{\dagger} - \frac{1}{2}\tilde V_{i}^{\dagger}\tilde V_i\tilde\rho - \frac{1}{2}\tilde \rho \tilde V_{i}\tilde V_{i}^{\dagger}\biggr\}
\end{equation}
with $\tilde V=UVU^{\dagger}$. The explicit form of Lindblad operator is determined by the type of environment to consider ~\cite{Suman} at zero temperature.  So, the operators can be    $V_i=S_i^-$ (for dissipation) for the model independent with the environment.  In this case, each qubit of the chain acts independently with the environment, and one has local decoherence of the system. The Lindblad operator is
\begin{equation}
\tilde{\mathcal{L}}(\tilde\rho)  =  \frac{1}{2i\hbar}\sum_{k}^{N}\gamma_{k}\bigl(2\tilde{S}_{k}^{-}\tilde\rho\tilde{S}_{k}^{+} - S_k^+S_k^-\tilde\rho - \tilde\rho \tilde S_k^+\tilde S_k^-\bigr)
\end{equation}
where $\tilde S_k^{+}$ and $\tilde S_k^{-}$ are the ascend and descend operators such that $ \tilde{S}_k^{\pm} = U S_k^{\pm}U^{\dagger} =S_k^{\pm}e^{\pm i\hat{\Omega}_k t}$,
where $\hat{\Omega}_k$ is defined as ${\hat{\Omega}_k = w_k + \frac{J}{\hbar}(S_{k+1}^z + S_{k-1}^z) + \frac{J'}{\hbar}(S_{k+2}^z + S_{k-2}^z).}$. The solutions of the equations are
\begin{eqnarray*}
 \rho_{11}(t) &=& \rho_{11}(0)+ \rho_{22}(0)+ \rho_{33}(0)+ \rho_{44}(0)+ \rho_{55}(0)+ \rho_{66}(0)+ \rho_{77}(0)+ \rho_{88}(0)\\
                         & & -( \rho_{55}(0)+ \rho_{66}(0)+ \rho_{77}(0)+ \rho_{88}(0))e^{-\gamma_1t}-( \rho_{33}(0)+ \rho_{44}(0)+ \rho_{77}(0)+ \rho_{88}(0))e^{-\gamma_2t}\\
                         & & -( \rho_{22}(0)+ \rho_{44}(0)+ \rho_{66}(0)+ \rho_{88}(0))e^{-\gamma_3t}+( \rho_{77}(0)+ \rho_{88}(0))e^{-(\gamma_1+\gamma_2)t}\\
                         & & +( \rho_{66}(0)+ \rho_{88}(0))e^{-(\gamma_1+\gamma_3)t}+( \rho_{44}(0)+ \rho_{88}(0))e^{-(\gamma_2+\gamma_3)t}- \rho_{88}(0)e^{-(\gamma_1+\gamma_2+\gamma_3)t};\\
 \rho_{14}(t) &=& \left[ \rho_{14}(0)+\frac{\gamma_1 \rho_{58}(0)e^{-i\phi_{14}}}{\sqrt{\gamma_1^2+(j+j')^2}}\left(1-e^{(i(j+j')-\gamma_1)t}\right)\right]e^{-\frac{1}{2}(\gamma_2+\gamma_3)t};\\
 \rho_{16}(t) &=& \left[ \rho_{16}(0)+\frac{\gamma_2 \rho_{38}(0)e^{-i\phi_{16}}}{\sqrt{\gamma_2^2+4j^2}}\left(1-e^{(2ij-\gamma_2)t}\right)\right]e^{-\frac{1}{2}(\gamma_1+\gamma_3)t};\\
 \rho_{17}(t) &=& \left[ \rho_{17}(0)+\frac{\gamma_3 \rho_{28}(0)e^{-i\phi_{17}}}{\sqrt{\gamma_3^2+(j+j')^2}}\left(1-e^{[i(j+j')-\gamma_3]t}\right)\right]e^{-\frac{1}{2}(\gamma_1+\gamma_2)t};\\
 \rho_{18}(t) &=&  \rho_{18}(0)e^{-\frac{1}{2}(\gamma_1+\gamma_2+\gamma_3)t};\\
\end{eqnarray*}

\begin{eqnarray*}
 \rho_{22}(t) &=& ( \rho_{22}(0)+ \rho_{44}(0)+ \rho_{66}(0)+ \rho_{88}(0))e^{-\gamma_3t}-( \rho_{66}(0)+ \rho_{88}(0))e^{-(\gamma_1+\gamma_3)t}\\
                         & &-( \rho_{44}(0)+ \rho_{88}(0))e^{-(\gamma_2+\gamma_3)t}+ \rho_{88}(0)e^{-(\gamma_1+\gamma_2+\gamma_3)t};\\
 \rho_{23}(t) &=& \left[ \rho_{23}(0)+\frac{\gamma_1 \rho_{67}(0)e^{-i\phi_{23}}}
{\sqrt{\gamma_1^2+(j-j')^2}}\left(1-e^{(i(j-j')-\gamma_1)t}\right)\right]e^{-\frac{1}{2}(\gamma_2+\gamma_3)t};\\
 \rho_{25}(t) &=& \left[ \rho_{25}(0)+ \rho_{47}(0)\left(1-e^{-\gamma_2t}\right)\right]e^{-\frac{1}{2}(\gamma_1+\gamma_3)t};\\
 \rho_{27}(t) &=&  \rho_{27}(0)e^{-\frac{1}{2}(\gamma_1+\gamma_2+\gamma_3)t};\\
 \rho_{28}(t) &=&  \rho_{28}(0)e^{-\frac{1}{2}(\gamma_1+\gamma_2+2\gamma_3)t};\\
 \rho_{33}(t) &=& ( \rho_{33}(0)+ \rho_{44}(0)+ \rho_{77}(0)+ \rho_{88}(0))e^{-\gamma_2t}-( \rho_{77}(0)+ \rho_{88}(0))e^{-(\gamma_1+\gamma_2)t}\\
                         & &-( \rho_{44}(0)+ \rho_{88}(0))e^{-(\gamma_2+\gamma_3)t}+ \rho_{88}(0)e^{-(\gamma_1+\gamma_2+\gamma_3)t};\\
 \rho_{35}(t) &=& \left[ \rho_{35}(0)+\frac{\gamma_3 \rho_{46}(0)e^{-i\phi_{35}}}
{\sqrt{\gamma_3^2+(j-j')^2}}\left(1-e^{-[i(j-j')+\gamma_3]t}\right)\right]e^{-\frac{1}{2}(\gamma_1+\gamma_2)t};\\
 \rho_{36}(t) &=&  \rho_{36}(0)e^{-\frac{1}{2}(\gamma_1+\gamma_2+\gamma_3)t};\\
 \rho_{38}(t) &=&  \rho_{38}(0)e^{-\frac{1}{2}(\gamma_1+2\gamma_2+\gamma_3)t};\\
 \rho_{44}(t) &=& ( \rho_{44}(0)+ \rho_{88}(0))e^{-(\gamma_2+\gamma_3)t}- \rho_{88}(0)e^{-(\gamma_1+\gamma_2+\gamma_3)t};\\
 \rho_{45}(t) &=&  \rho_{45}(0)e^{-\frac{1}{2}(\gamma_1+\gamma_2+\gamma_3)t};\\
 \rho_{46}(t) &=&  \rho_{46}(0)e^{-\frac{1}{2}(\gamma_1+\gamma_2+2\gamma_3)t};\\
 \rho_{47}(t) &=&  \rho_{47}(0)e^{-\frac{1}{2}(\gamma_1+2\gamma_2+\gamma_3)t};\\
 \rho_{48}(t) &=&  \rho_{48}(0)e^{-\frac{1}{2}(\gamma_1+2\gamma_2+2\gamma_3)t};\\
  \rho_{55}(t) &=& ( \rho_{55}(0)+ \rho_{66}(0)+ \rho_{77}(0)+ \rho_{88}(0))e^{-\gamma_1t}-( \rho_{77}(0)+ \rho_{88}(0))e^{-(\gamma_1+\gamma_2)t}\\
                         & &-( \rho_{66}(0)+ \rho_{88}(0))e^{-(\gamma_1+\gamma_3)t}+ \rho_{88}(0)e^{-(\gamma_1+\gamma_2+\gamma_3)t};\\
 \rho_{58}(t) &=&  \rho_{58}(0)e^{-\frac{1}{2}(2\gamma_1+\gamma_2+\gamma_3)t};\\
 \rho_{66}(t) &=& ( \rho_{66}(0)+ \rho_{88}(0))e^{-(\gamma_1+\gamma_3)t}- \rho_{88}(0)e^{-(\gamma_1+\gamma_2+\gamma_3)t};\\
 \rho_{67}(t) &=&  \rho_{67}(0)e^{-\frac{1}{2}(2\gamma_1+\gamma_2+\gamma_3)t};\\
 \rho_{68}(t) &=&  \rho_{68}(0)e^{-\frac{1}{2}(2\gamma_1+\gamma_2+2\gamma_3)t};\\
 \rho_{77}(t) &=& ( \rho_{77}(0)+ \rho_{88}(0))e^{-(\gamma_1+\gamma_2)t}- \rho_{88}(0)e^{-(\gamma_1+\gamma_2+\gamma_3)t};\\
 \rho_{78}(t) &=&  \rho_{78}(0)e^{-\frac{1}{2}(2\gamma_1+2\gamma_2+\gamma_3)t};\\
 \rho_{88}(t) &=&  \rho_{88}(0)e^{-(\gamma_1+\gamma_2+\gamma_3)t}.
\end{eqnarray*}
where $\phi_{ij}$ are given by
\begin{align*}
 \phi_{14} & = \tan^{-1}\left(\frac{j+j'}{\gamma_1}\right);   & \phi_{16} & = \tan^{-1}\left(\frac{2j}{\gamma_2}\right);    & \phi_{17} & = \tan^{-1}\left(\frac{j+j'}{\gamma_3}\right);\\
 \phi_{23} & = \tan^{-1}\left(\frac{j-j'}{\gamma_1}\right);   & \phi_{35} & = \tan^{-1}\left(\frac{j-j'}{\gamma_3}\right).  &           & 
\end{align*}
\\
In our case, we have three qubits space $\{|\xi_3\xi_2\xi_1\rangle\}_{\xi_i=0,1}$, and our parameter in units $2\pi MHz$ are
\begin{eqnarray*}
\omega_1    & = & 400; \quad \omega_2     = 200;  \quad \omega_3    = 100    \quad J          = 10;   \quad J'          = 0.4      \\
\gamma_1    & = & 0.05;\quad \gamma_2     = 0.05;  \quad \gamma_3    = 0.05.
 \end{eqnarray*}
the time is normalized by the same factor of $2\pi MHz$, and we include in this study the entangle state
\begin{equation}
|\Psi_1\rangle=\frac{1}{2}\bigl(|000\rangle+|111\rangle+|001\rangle+|110\rangle\bigr).
\end{equation}
Figure 4 shows the behavior of the entangled states $|W\rangle$ and $|GHZ\rangle$ as a function of time when this entangled state interact with the environment. Purity behavior, $tr(\rho^2)$, is also shown. The system starts as a pure entangled state, it evolves in a mixed state and finishes in the pure ground state ($|000\rangle$), after sharing energy with the environment.  The states $|GHZ\rangle$ and $|\Psi_1\rangle$ behave more robust than the state $|W\rangle$, $C_{\rho}^{(3)}$ grows since other entangled stated contribute to this function. In contrast, starting with the entangled state $|W\rangle$, there are not other entangled states which make contribution to the function $C_{\rho}^{(3)}$ in the dynamics, and one sees an exponential decay. 
\noindent
\begin{figure}[H]
\includegraphics[scale=0.60,angle=0]{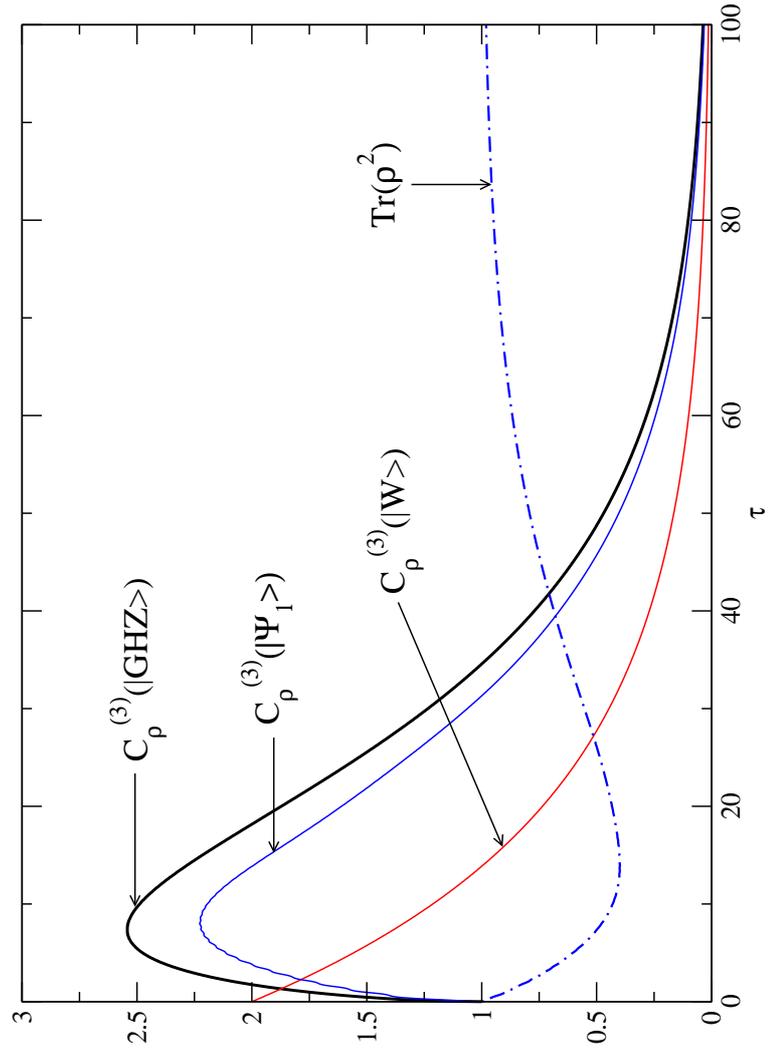}
\centering
    \caption{ $C_{\rho}^{(3)}$ and Purity for the entangled state $|W\rangle$,  $|GHZ\rangle$, and $|\Psi_1\rangle$. }
\end{figure}
\section{Conclusions}
We have studied the full factorization of an state made up of up to 4-qubits basic states. We have seen that there is an indication that the number of conditions to characterizes a factorized state grows exponentially with the number of qubits.  For two, three and four basic qubits, we showed the conditions in order to have a factorized state, and if any of one of this conditions fails, one gets instead an entangled state. Therefore, an entangled state is also characterized by the complement of each of this conditions, and the resulting expression has been denoted by $C^{(n)}$ (n=2,3,4). This non negative function expressed in terms of the coefficients of the wave function or in terms the density matrix elements  represents a measurement of the entanglement of any wave function made up of basic n-qubits ($n=2,3,4$). Using this function, we study the decay of entangled states $|W\rangle$,  $|GHZ\rangle$, and $|\Psi_1\rangle$ due to interaction with environment, and we noticed a great different behavior of the function $C^{(3)}_{\rho}$, indicating some type of robustness behavior of the states $|GHZ\rangle$ and $|\Psi_1\rangle$. The main reason for this different behavior  is that the entangled states $|GHZ\rangle$ and $|\Psi_1\rangle$ contain the ground state $|000\rangle$, which is the final state in the dynamics.     

\begin{thebibliography}{99}
%
   \bibitem{Sakurai}  Sakurai J. J. {Modern Quantum Mechanics (Revised Edition)} Addison Wesley; 1 edition ( 1993).
   \bibitem{Mess} A. Messiah, {\it Quantum Mechanics Vol. II}, John-Wiley and Sons, (1958).
   \bibitem{Sch} Schr\"{o}dinger E., {\it Naturwissenschaften} 23 807, 823, 844 (1935).
  \bibitem{Horo} Horodecki R., Horodecki P., Horodecki M., Horodecki K., {Rev. Mod. Phys.} {\bf 81}, 865 (2009).
   \bibitem{Woott} Wootters W. K., {Phys. Rev. Lett.} {\bf 80}, 2245 (1998).
   \bibitem{Seev}M. Seevinck and J. Uffink, {\it Partial separability and entanglement criteria for multiqubit quantum states}, {Phys. Rev.} A {\bf 78}, 032101 (2008).
   \bibitem{Aoli} Aolita L., Chaves R., Cavalcanti D., Ac\'in A., Davidovich L., {Phys. Rev. Lett.} {\bf 100}, 080501 (2008).
   \bibitem{Pope} Pope D. T., Milburn G. J., {Phys. Rev.} A {\bf 67}, 052107 (2003).
   \bibitem{Love} Love P., van den Brink A., Smirnov A., Amin A., Gra-jcar M., Ilichev E., Izmalkov A., Zagoskin A., {Quant. Inf. Proc.} {\bf 6}, 187 (2007).
   \bibitem{Zhi} Zhi-Hao Ma, Zhi-Hua Chen, and Jing-Ling Chen, {\it  Measure of genuine multipartite entanglement with computable lower bounds }, {Phys. Rev.} A {\bf 83}, 062325 (2011).
   \bibitem{Huber}M. Huber and F.  Mintert Detection of High-Dimensional Genuine Multipartite Entanglement of Mixed States {Phys. Rev. Lett.} {\bf 104}, 210501 (2010).
   \bibitem{Lin} L.Chen and Yi-Xin Chen, {\it Phys. Rev. A}, {\bf 76}, 022330 (2007).
   \bibitem{Ry} R. Horodecki, P. Horodecki, M. Horodecki, and K. Horodecki, {\it Quantum Entanglement}, {\it Rev. Mod. Phys}. {\bf 81}, 865 (2009). 
   \bibitem{Dur} W. D\"ur and J.I. Chirac, {\it Classification of multiqubit mixed states: Separability and distillability properties}, {\it Phys. Rev. A}, {\bf 61}, 042314 (2000).
    \bibitem{Niel_QC}Nielsen M and Chuang I \textit{Quantum Computation and Quantum Information} (Cambridge: Cambridge University Press) (2004).
    \bibitem{Ko} A.N. Kolmogorov and S.V. Fomin, {\it Introductory Real Analysis}, Dover Publications Inc., (1970).
   \bibitem{Lloyd_93} S. Lloyd, {\it A potential Realizable Quantum Computer}, Science, {\bf 261},  1569 (1993).
   \bibitem{Lopez1} G.V. L\'opez, {\it Diamond as a Solid State Quantum Computer with a Linear Chain of Nuclear Spins System}, {\it J. Mod. Phys., } {\bf  5}, 55 (2014).
    \bibitem{Berman} Berman G.P., Doolen D.D., Kamenev D.I., L\'opez G.V., Tsifrinovich  V.I., {\it Perturbation Theory and Numeri-cal Modeling of     Quantum Logic Operations with Large Number of Qubits},  {Contemp. Math.} {\bf 305}, pp.13-41 (2002).   
   \bibitem{Fano}U. Fano, {\it Description of States in Quantum Mechanics by Density Matrix and Operator Techniques}, Rev. Mod. Phys., {\bf 29}, 74 (1957).  
   \bibitem{von} J. von Neumann, {\it Wahrsheinlichkeitstheoretischer Aufbau der Quantenmechanik}, G\"ottinger Nachrichten, {\bf 1}, 245 (1927).   
    \bibitem{Davies}E.B. Davies, {\it Quantum Theory of Open Systems}. Academic Press, San Diego (1976).
    \bibitem{Bre_OQS}Breuer, Heinz-Peter; F. Petruccione, {\it The Theory of Open Quantum Systems}. Oxford University Press (2007). 
    \bibitem{Alicki}R. Alicki, K. Lendi, {\it Quantum Dynamical Semigroups and Applications}. Lecture Notes Phys., vol. 717. Springer, Berlin (2007).
    \bibitem{Suman}Das Sumanta  and G.S. Agarwal, {\it Decoherence effects in interacting qubits under the influence of various environments}, J. Phys. B: At. Mol.  Opt. Phys., {\bf 42} (2009).
    \bibitem{Alb} S. Alberio and Shao-Ming Fei, {\it A note on Invariants and Entanglements}, quant-ph/0109073v1, 17 Sep (2001).
   
%
  \end{thebibliography}
\end{document}